\documentclass[a4paper,12pt]{article}

\usepackage{latexsym,amsmath,amsfonts,amssymb}
\usepackage{multirow}
\usepackage{float}
\usepackage{graphicx,color}
\usepackage{subfigure}
\usepackage{mathrsfs,mathtools}
\usepackage{pstricks,pst-node,pst-text,pst-3d}
\usepackage{verbatim}
\usepackage{epsfig}

\newcommand{\sect}[1]{\setcounter{equation}{0}\section{#1}}

\begin{document}

\author{Francis Bursa}
\author{Luigi Del Debbio}
\author{Liam Keegan}
\author{Claudio Pica}
\author{Thomas Pickup}
\title{Mass anomalous dimension in SU(2) with two adjoint fermions}

\begin{titlepage}

\setcounter{page}{0}

\begin{flushright}
DAMTP-2009-65
\end{flushright}

\vspace{0.6cm}

\begin{center}
{\Large \bf Mass anomalous dimension in SU(2) with two adjoint fermions.} \\ 

\vskip 0.8cm

{\bf Francis Bursa}\\
{\sl
Jesus College, Cambridge, CB5 8BL, United Kingdom}\\
{\bf Luigi Del Debbio, Liam Keegan, Claudio Pica}\\
{\sl
SUPA, School of Physics and Astronomy, \\
University of Edinburgh \\
Edinburgh EH9 3JZ, United Kingdom}\\
{\bf Thomas Pickup}\\
{\sl Rudolf Peierls Centre for Theoretical Physics,\\
 University of Oxford, Oxford OX1 3NP, United Kingdom}

\vskip 1.2cm

\end{center}

\begin{abstract}
  We study $SU(2)$ lattice gauge theory with two flavours of Dirac
  fermions in the adjoint representation. We measure the running of
  the coupling in the Schr\"{o}dinger Functional (SF) scheme and find
  it is consistent with existing results.  We discuss how systematic
  errors affect the evidence for an infrared fixed point (IRFP). We
  present the first measurement of the running of the mass in the SF
  scheme. The anomalous dimension of the chiral condensate, which is
  relevant for phenomenological applications, can be easily extracted
  from the running of the mass, under the assumption that the theory
  has an IRFP.  At the current level of accuracy, we can estimate
  $0.05 < \gamma < 0.56$ at the IRFP.
\end{abstract}

\vfill

\end{titlepage}

\sect{Introduction}
\label{sec:intro}
Experiments at the LHC are about to probe nature at the TeV scale,
where new physics beyond the Standard Model (BSM) is expected to be
found. The existence of a new strongly--interacting sector that is
responsible for electroweak symmetry breaking is an interesting
possibility. Technicolor was originally proposed thirty years ago, and
strongly--interacting BSM has been revisited in many instances since
then. Recent reviews can be found in
Refs.~\cite{Hill:2002ap,Sannino:2008ha}.

In order to be phenomenologically viable, technicolor theories need to
obey the constraints from precision measurements at
LEP~\cite{Peskin:1990zt,Altarelli:1990zd}. Moreover the symmetry
breaking needs to be communicated to the Standard Model, so that the
usual low--energy physics is recovered. This is usually achieved in
the so-called Extended Technicolor (ETC) models by invoking some
further interaction at higher energies that couples the technicolor
sector to the Standard Model. At the TeV scale the remnants of this
coupling are higher--dimensional operators in the effective
Hamiltonian, which are suppressed by powers of the high energy scale,
$M$, that characterises the extended model. Amongst these operators
are a mass term for the Standard Model quarks, and four--fermion
interactions that would contribute to flavour--changing neutral
currents (FCNC). Thus there is a tension on the possible values of
$M$: on the one hand $M$ needs to be large so that FCNC interactions
are suppressed, on the other hand $M$ needs to be small enough to
generate the heavier quark masses. In particular, the effective
operator for the Standard Model quark masses is:
\begin{equation}
  \label{eq:qmass}
  \mathcal L_m = \frac{1}{M^2} \langle \Phi \rangle \bar\psi \psi\, ,
\end{equation}
where $\psi$ indicates the quark field, and $\Phi$ is the field in the
technicolor theory which is responsible for electroweak symmetry
breaking. In the traditional technicolor models, which are realised as
SU($N$) gauge theories, $\Phi=\bar\Psi \Psi$ is the chiral condensate
of techniquarks. Let us emphasise that quark masses are defined in a
given renormalisation scheme and at a given scale. For instance the
data reported in the Particle Data Group summaries~\cite{pdg:2009}
usually refer to the quark mass in the $\overline{\mathrm{MS}}$ scheme
at 2 GeV. The coefficient that appears in Eq.~(\ref{eq:qmass}) is the
chiral condensate at the scale $M$:
\begin{equation}
  \label{eq:chrun}
  \left. \langle \bar\Psi\Psi \rangle \right|_M = 
  \left. \langle \bar\Psi\Psi \rangle \right|_\Lambda 
  \exp\left[\int_\Lambda^M \frac{d\mu}{\mu} \gamma(\mu)\right]\, ,
\end{equation}
where $\gamma$ is the anomalous dimension of the scalar density,
and $\Lambda$ is the typical scale of the technicolor theory, 
$\Lambda\approx 1~\mathrm{TeV}$. The chiral condensate at this scale
is expected to be $\langle\bar\Psi\Psi\rangle\sim\Lambda^3$, and
therefore the naive expectation for the quark masses is $m\sim
\Lambda^3/M^2$.

Eq.~(\ref{eq:chrun}) suggests a possible way to resolve the tension
due to the large quark masses. If the technicolor theory is such that
$\gamma$ is approximately constant (and large) over a sufficiently
long range in energies, then the running above will generate a power
enhancement of the condensate. This scenario has been known for a long
time under the name of {\it walking
technicolor}~\cite{Holdom:1984sk,Holdom:1983kw,Yamawaki:1985zg}. Gauge
theories with a large number of fermions have been traditional
candidates for walking theories; the fermions slow down the running of
the coupling and can potentially lead to the required
power--enhancement. More recent incarnations have been proposed that
are constructed as SU(N) gauge theories with fermions in
higher--dimensional representations of the colour
group~\cite{Dietrich:2006cm,Foadi:2007ue,Foadi:2007se}. These theories
could have a genuine IR fixed point (IRFP), or simply lie in its
vicinity. The existence of an IRFP is a difficult problem to address
since it requires to perform quantitative computations in a
strongly--interacting theory. Lattice simulations can provide
first--principle results that can help in determining the
phenomenological viability of these models; numerical simulations of
models of dynamical electroweak symmetry breaking have attracted
growing attention in recent
years~\cite{Catterall:2007yx,Appelquist:2007hu,DelDebbio:2008wb,Shamir:2008pb,Deuzeman:2008sc,DelDebbio:2008zf,Catterall:2008qk,Svetitsky:2008bw,DeGrand:2008dh,Fodor:2008hm,Fodor:2008hn,Deuzeman:2008pf,Deuzeman:2008da,Hietanen:2008vc,Jin:2008rc,DelDebbio:2008tv,DeGrand:2008kx,Fleming:2008gy,Hietanen:2008mr,Appelquist:2009ty,Hietanen:2009az,Deuzeman:2009mh,Fodor:2009nh,DeGrand:2009mt,DeGrand:2009et,Hasenfratz:2009ea,DelDebbio:2009fd,Fodor:2009wk,Pica:2009hc}. A
number of theories have been studied: SU(3) with 8, 10, 12 flavours of
fermions in the fundamental representation, SU(3) with fermions in the
sextet representation, and SU(2) with fermions in the adjoint
representation. These studies have focused either on the spectrum of
the theories, or on the running of the coupling computed in the
Schr\"odinger functional (SF) scheme, finding some tantalising
numerical evidence for IR behaviours different from what is known from
QCD.

Existing simulations of the Schr\"odinger functional have identified a
possible fixed point in many of the above--mentioned theories by
noticing a flat behaviour of the running coupling in this scheme over
a given range of energy scales.

In this work we consider the SU(2) theory with two flavours of adjoint
fermions, and compute the running coupling in the SF scheme. We
confirm the results obtained in Ref.~\cite{Hietanen:2009az}, and
present a more refined analysis of the lattice data. We focus on the
running of the mass in the SF scheme, from which we can extract the
mass anomalous dimension that appears in Eq.~(\ref{eq:chrun}). Current
simulations are still plagued by systematic errors, which we examine
in detail both for the coupling and the mass. These errors are the
largest limitation to drawing strong conclusions from the lattice
data. These limitations are common to all the studies performed so
far, more extensive work is required in order to reach robust
conclusions. Our results for the anomalous dimension of the mass
provide crucial input for these studies that aim at exact results for
non-supersymmetric gauge theories in the non-perturbative regime.

\sect{SF formulation}
\label{sec:SF}
\subsection{Basic definitions}

We define the running coupling $\overline{g}^2$ non-perturbatively
using the Schr\"{o}dinger Functional
scheme~\cite{Luscher:1991wu,Luscher:1992an}. This is defined on a
hypercubic lattice of size $L$, with boundary conditions chosen to
impose a background chromoelectric field on the system. The
renormalised coupling is defined as a measure of the response of the
system to a small change in the background chromoelectric
field. Specifically, the spatial link matrices at $t=0$ and $t=L$ are
set respectively to:
\begin{eqnarray}
  \label{eq:linkBC1}
  \left.U(x,k)\right|_{t=0}&=&\exp\left[\eta \tau_3 a/iL\right]\, , \\
  \label{eq:linkBC2}
  \left.U(x,k)\right|_{t=L}&=&\exp\left[(\pi-\eta) \tau_3a/iL\right]\, ,
\end{eqnarray}
with $\eta=\pi/4$~\cite{Luscher:1992zx}. The fermion fields obey 
\begin{eqnarray}
  \label{eq:ferm0}
  P_+\psi=0,~\overline{\psi}P_-=0&&~~\mathrm{at}~~t=0\, , \\
  \label{eq:fermL}
  P_-\psi=0,~\overline{\psi}P_+=0&&~~\mathrm{at}~~t=L\, ,
\end{eqnarray}
where the projectors are defined as $P_\pm=(1\pm\gamma_0)/2$. The
fermion fields also satisfy periodic spatial boundary
conditions~\cite{Sint:1995ch}. As we mentioned above, one can readily
verify in perturbation theory that these boundary conditions impose a
constant chromoelectric field.

We use the Wilson plaquette gauge action, and Wilson fermions in the
adjoint representation, as implemented in
Ref.~\cite{DelDebbio:2008zf}. Note that we have not improved the
action, and therefore our results are going to be affected by $O(a)$
lattice artefacts. The same approach has been used so far for the
preliminary studies of this theory in Ref.~\cite{Hietanen:2009az}.

The coupling constant is defined as
\begin{equation}
  \label{eq:SFcoupling}
  \overline{g}^2=k \left< \frac{\partial S}{\partial \eta} \right>^{-1}
\end{equation}
with $k=-24L^2/a^2 \mathrm{sin}(a^2/L^2 (\pi-2\eta))$ chosen such that
$\overline{g}^2=g_0^2$ to leading order in perturbation theory. This
gives a non--perturbative definition of the coupling which depends on
only one scale, the size of the system $L$.

To measure the running of the quark mass, we calculate the
pseudoscalar density renormalisation constant $Z_P$. Following
Ref.~\cite{Capitani:1998mq}, $Z_P$ is defined by:
\begin{equation}
  \label{eq:ZPdef}
  Z_P(L)=\sqrt{3 f_1}/f_P(L/2)\, ,
\end{equation}
where $f_1$ and $f_P$ are the correlation functions involving the
boundary fermion fields $\zeta$ and $\overline{\zeta}$:
\begin{eqnarray}
  \label{eq:f1def}
  f_1&=&-1/12L^6 \int d^3u\, d^3v\, d^3y\, d^3z\,
  \langle
  \overline{\zeta}^\prime(u)\gamma_5\tau^a{\zeta}^\prime(v)
  \overline{\zeta}(y)\gamma_5\tau^a\zeta(z) 
  \rangle\, , \\
  \label{eq:fPdef}
  f_P(x_0)&=&-1/12 \int d^3y\, d^3z\,  \langle
  \overline{\psi}(x_0)\gamma_5\tau^a\psi(x_0)\overline{\zeta}(y)\gamma_5\tau^a\zeta(z)
  \rangle\, .
\end{eqnarray}
These correlators are calculated on lattices of size $L$, with the
spatial link matrices at $t=0$ and $t=L$ set to unity.

The Schr\"{o}dinger Functional boundary conditions remove the zero
modes that are normally an obstacle to simulating at zero quark
mass~\cite{Sint:1993un}. This means we can run directly at $\kappa_c$.
We determine $\kappa_c$ through the PCAC mass in units of the inverse
lattice spacing $am(L/2)$, where
\begin{equation}
am(x_0)=\frac{\frac{1}{2}(\partial_0+\partial_0^*)f_A(x_0)}{2f_P(x_0)}
\end{equation}
and
\begin{equation}
  f_A(x_0)=-1/12 \int d^3yd^3z \langle
  \overline{\psi}(x_0)\gamma_0\gamma_5\tau^a\psi(x_0)\overline{\zeta}(y)\gamma_5\tau^a\zeta(z) \rangle.
\end{equation}
Here the lattice derivatives $\partial_0$ and $\partial_0^*$ are
defined by $\partial_0f(x)=f(x+1)-f(x)$ and
$\partial_0^*f(x)=f(x)-f(x-1)$, and the correlators are calculated on
lattices of size $L$, with the spatial link matrices at $t=0$ and
$t=L$ set to unity.

We define $\kappa_c$ by the point where $am$ vanishes. We measure $am$
for 5 values of $\kappa$ in the region $-0.2< am <0.2$ and use a
linear interpolation in $\kappa$ to find an estimate of
$\kappa_c$. The error on $\kappa_c$ is estimated by the bootstrap
method. 

In practice we achieve $|am|\lesssim 0.005$.  We check explicitly that
there is no residual sensitivity to the small remaining quark mass by
repeating some of our simulations at moderately small values of $am
\sim 0.02$, for which we found no shift in $\overline{g}^2$ or $Z_P$
within the statistical uncertainty of the measured values, so the
effect of our quark mass can safely be neglected.

\subsection{Lattice parameters}
We have performed two sets of simulations in order to determine the
running coupling and $Z_P$. The parameters of the runs are summarised
respectively in Tab.~\ref{tab:gpar}, and~\ref{tab:ZPpar}. The values
of $\kappa_c$ are obtained from the PCAC relation as described above.

\begin{table}[H]
  \centering

{\footnotesize
  \begin{tabular}{|c|c|c|c|c|}
    \hline
    $\beta$ & $L$=6 & $L$=8 & $L$=12 & $L$=16\\
    \hline
    2.00 & 0.190834 & - & - & - \\
    2.10 & 0.186174 & - & - & - \\
    2.20 & 0.182120 & 0.181447 & 0.1805 & - \\
    2.25 & 0.180514 & 0.179679 & - & - \\
    2.30 & 0.178805 & 0.178045 & - & - \\
    2.40 & 0.175480 & 0.174887 & - & - \\
    2.50 & 0.172830 & 0.172305 & 0.17172 & 0.17172 \\
    2.60 & 0.170162 & 0.169756 & - & - \\
    2.70 & 0.167706 & - & - & - \\
    \hline
  \end{tabular}
  \begin{tabular}{|c|c|c|c|c|}
    \hline
    $\beta$ & $L$=6 & $L$=8 & $L$=12 & $L$=16\\
    \hline
    2.80 & 0.165932 & 0.165550 & 0.16505 & - \\
    3.00 & 0.162320 & 0.162020 & 0.161636 & 0.161636 \\
    3.25 & 0.158505 & - & 0.1580 & - \\
    3.50 & 0.155571 & 0.155361 & 0.155132 & 0.155132 \\
    3.75 & 0.152803 & - & - & - \\
    4.00 & 0.150822 & 0.150655 & - & - \\
    4.50 & 0.147250 & 0.14720 & 0.14712 & 0.14712 \\
    8.00 & 0.136500 & 0.13645 & 0.136415 & - \\    
    & & & & \\
    \hline
  \end{tabular}
}

\caption{Values of $\beta$, $L$, $\kappa$ used for the determination of
  $\overline{g}^2$. The entries in the table are the values 
  of $\kappa_c$ used for each
  combination of $\beta$ and $L$. 
}

\label{tab:gpar}
\end{table}

Note that $Z_P$ is determined from a different set of runs at similar
values of $\beta$, $L$, $\kappa$. 

\begin{table}[H]
  \centering

 {\footnotesize
  \begin{tabular}{|c|c|c|c|c|}
    \hline
    $\beta$ & $L$=6 & $L$=8 & $L$=12 & $L$=16\\
    \hline
    2.00 & 0.190834 & - & - & - \\
    2.05 & 0.188504 & - & 0.18625 & - \\
    2.10 & 0.186174 & - & - & - \\
    2.20 & 0.182120 & 0.181447 & 0.1805 & - \\
    2.25 & 0.180514 & 0.179679 & - & - \\
    2.30 & 0.178805 & 0.178045 & - & - \\
    2.40 & - & 0.174887 & - & - \\
    2.50 & 0.172830 & 0.172305 & 0.17172 & 0.17172 \\
    2.60 & 0.170162 & 0.169756 & - & - \\
    2.70 & 0.167706 & - & - & - \\
    \hline
  \end{tabular}
  \begin{tabular}{|c|c|c|c|c|}
    \hline
    $\beta$ & $L$=6 & $L$=8 & $L$=12 & $L$=16\\
    \hline
    2.80 & 0.165932 & 0.165550 & 0.16505 & - \\
    3.00 & 0.162320 & 0.162020 & 0.161636 & 0.161636 \\
    3.25 & 0.158505 & - & 0.1580 & - \\
    3.50 & 0.155571 & 0.155361 & 0.155132 & 0.155132 \\
    3.75 & 0.152803 & - & - & - \\
    4.00 & 0.150822 & 0.150655 & 0.15051 & - \\
    4.50 & 0.14725 & 0.14720 & 0.14712 & 0.14712 \\
    8.00 & 0.13650 & 0.13645 & 0.136415 & 0.136415 \\
    16.0 & 0.1302 & 0.1302 & 0.1302 & 0.130375 \\
    & & & & \\
    \hline
  \end{tabular}
}

\caption{Values of $\beta$, $L$, $\kappa$ used for the determination of
  $Z_P$. The entries in the table are the values of $\kappa_c$ used for each
  combination of $\beta$ and $L$. 
}
\label{tab:ZPpar}
\end{table}

\sect{Evidence for fixed points}
\label{sec:scheme}
Recent studies have focused on the running of the SF gauge coupling,
and have highlighted a slow running in the lattice data for this
quantity~\cite{Appelquist:2007hu,Shamir:2008pb,Appelquist:2009ty,Hietanen:2009az}. This
is clearly different from the behaviour observed in QCD--like
theories~\cite{Luscher:1992zx,DellaMorte:2004bc}. These results are
certainly encouraging, but have to be interpreted with care. Lattice
data can single out at best a range of energies over which no running
is observed. However it is not possible to conclude from lattice data
only that the plateau in the running coupling does extend to
arbitrarily large distances, as one would expect in the presence of a
genuine IRFP. On the other hand, if the plateau has a finite extent,
{\it i.e.} if the theory seems to walk only over a finite range of
energies, then the behaviour of the running coupling in the absence of
a genuine fixed point depends on the choice of the scheme, and
therefore the conclusions become less compelling.

Let us discuss the scheme dependence of the running coupling in more
detail. The quantities we are interested in are the beta function and
the mass anomalous dimension:
\begin{eqnarray}
  \label{eq:betaf}
  \mu \frac{d}{d\mu} \overline{g}(\mu) &=& \beta(\overline{g}) \, , \\
  \label{eq:gammaf}
  \mu \frac{d}{d\mu} \overline{m}(\mu) &=& -\gamma(\overline{g}) \overline{m}(\mu) \, ,
\end{eqnarray}
where $\overline{g},\overline{m}$ are the running coupling and mass in
a given (mass-independent) renormalisation scheme. Note that $\gamma$
in Eq.~(\ref{eq:gammaf}) is the anomalous dimension of the scalar
density, which appears also in Eq.~(\ref{eq:chrun}); $\gamma$ differs
from the usual mass anomalous dimension by an overall sign. Both
$\beta$ and $\gamma$ can be computed in perturbation theory for small
values of the coupling constant:
\begin{eqnarray}
  \label{eq:betapert}
  \beta(\overline{g}) &=& -\overline{g}^3\left[\beta_0 + \beta_1 \overline{g}^2 
  + \beta_2 \overline{g}^4 + O(\overline{g}^6)\right] \, ,\\
  \label{eq:gammmapert}
  \gamma(\overline{g}) &=& \overline{g}^2\left[d_0 + d_1 \overline{g}^2 
  + O(\overline{g}^4)\right] \, .
\end{eqnarray}
The coefficient $\beta_0,\beta_1,d_0$ are scheme--independent;
expressions for $\beta_0,\beta_1$ for fermions in arbitrary
representations of the gauge group have been given in
Ref.~\cite{DelDebbio:2008wb}, while for the first coefficient of the
anomalous dimension, we have:
\begin{equation}
  \label{eq:doexpr}
  d_0=\frac{6 C_2(R)}{(4\pi)^2}\, ,
\end{equation}
where $C_2(R)$ is the quadratic Casimir of the fermions' colour
representation. In the specific case we are studying in this work
$d_0=3/(4\pi^2)$. 

Different schemes are related by finite renormalisations;
the running of the couplings in going from one scheme to the other is
readily obtained by computing the scale dependence with the aid of the
chain rule. Let us consider a change of scheme:
\begin{eqnarray}
  \label{eq:schemeg}
  \bar g^\prime &=& \phi(\bar g,m/\mu)\, , \\
  \label{eq:schemem}
  \bar m^\prime &=& \bar m\, \mathcal F(\bar g,\bar m/\mu)\, .
\end{eqnarray}
We impose two conditions on $\phi$: it must be invertible, and should
reduce to $\phi(\bar g) = \bar g + O(\bar g^3)$ for small values of
$\bar g$. Eq.~(\ref{eq:schemem}) encodes the fact that a massless
theory remains massless in any scheme. The picture simplifies
considerably if one considers only mass--independent renormalisation
schemes; the functions $\phi$ and $\mathcal F$ only depend on the
coupling $\bar g$, and one finds:
\begin{eqnarray}
  \label{eq:betap}
  \beta^\prime(\bar g^\prime) &=& \beta(\bar g)
  \frac{\partial}{\partial \bar g} \phi(\bar g) \, \\
  \label{eq:gammap}
  \gamma^\prime(\bar g^\prime) &=& \gamma(\bar g) + \beta(\bar g)
  \frac{\partial}{\partial \bar g} \log \mathcal
  F(\bar g) \, .
\end{eqnarray}
The scheme--independence of the coefficients $\beta_0,\beta_1,d_0$ can
be obtained by expanding the functions that describe the mapping
between the two schemes, $\phi$ and $\mathcal F$, in powers of
$\overline{g}^2$. Eqs.~(\ref{eq:betap}), (\ref{eq:gammap}) summarise
the main features that we want to highlight here. The conditions we
imposed on $\phi$ imply that $\frac{\partial}{\partial \bar g}
\phi(\bar g)>0$,
{\it i.e.}  asymptotic freedom cannot be undone by a change of
scheme. The existence of a fixed point is clearly scheme--independent:
if $\beta(\bar g^*)=0$ for some value $\bar g^*$ of the coupling, then
$\beta^\prime$ has also a zero. Note that the value of the critical
coupling changes from one scheme to the other, $\bar g^{\prime
*}=\phi(\bar g^*)$, however the existence of the fixed point is invariant.
Similarly, the anomalous dimension is scheme--independent at a fixed
point, since the second term in Eq.~(\ref{eq:gammap}) vanishes
there. Moreover, if the change of scheme only involves a redefinition
of the coupling, but leaves the mass unchanged, then the anomalous
dimension does not vary.

Unfortunately none of these conclusions holds in the absence of a
fixed point. In particular, a flat behaviour of the running coupling
over a finite range of energies can be obtained in any theory by a
suitably--chosen change of scheme.

It is worth stressing here another important point concerning the
numerical studies of running couplings. There are instances where the
beta function of an asymptotically free theory remains numerically
small. This is the case of the theory considered in this work, namely
SU(2) with 2 flavours of adjoint Dirac fermions, in the perturbative
regime. In this case the running of the coupling is very slow from the
very beginning, and this is independent of the possible existence of
an IRFP at larger values of the coupling. As a consequence high
numerical accuracy is needed in order to resolve a ``slow'' running;
therefore numerical studies of potential IRFP need high statistics,
and a robust control of systematics. In particular it is important to
extrapolate the step-scaling functions computed on the lattice to the
continuum limit, in order to eliminate lattice artefacts which could
bias the analysis of the dependence of the running coupling on the
scale. This is particularly relevant for the studies of potential
IRFP, since lattice artefacts could more easily obscure the small
running that we are trying to resolve. Some of these difficulties were
already noted in Ref.~\cite{Hietanen:2009az}; current results,
including the ones presented in this work, are affected by these
systematics.

More extensive simulations are therefore needed in order to remove the
lattice artefacts by performing a controlled extrapolation of the
lattice step scaling functions defined below in
Sect.~\ref{sec:SFcoupling}. The scale $L$ at which the coupling is
computed and the lattice spacing $a$ must be well separated. This last
step is a crucial ingredient in the SF scheme, since it decouples the
details of the lattice discretization from the running of the
couplings at the scale $L$ that we want to determine. Asymptotically
free theories are effectively described by a perturbative expansion at
small distances. In this regime, the degrees of freedom are the
elementary fermions and the gauge bosons, renormalized couplings can
be computed in perturbation theory, and different schemes can be
related by perturbative calculations. The evolution of the running
coupling can be followed starting from this high--energy regime and
moving towards larger distances. If the theory has an IRFP, the value
of the running coupling approaches some finite limit $\bar g^*$ as $L$
is increased, i.e. the running coupling {\em must} lie in the interval
$\left[0,\bar g^*\right]$. Its running can be traced from the UV
regime up to the limiting value, which is approached from
below. Larger values of $\bar g$ can be obtained in a lattice
simulation; however the interpretation of these points is less
transparent. One possibility is that the lattice theory in some region
of bare parameter space lies in the basin of attraction of some
non-trivial UV fixed point where a {\em different} continuum theory
can be defined. The running coupling would then approach the IRFP
value from above. The non-trivial UV fixed point is clearly difficult
to identify, thereby making the extrapolation to the continuum limit
rather tricky in this case. 

A more pragmatic approach could be to ignore the issue of the
existence of a non-trivial UV fixed point, and simply explore the
limit $L/a\gg 1$, assuming that the starting point is the lattice
theory with a cutoff, and that we are only interested in the regime
where distances are large compared to the cutoff. This interpretation
is prone to systematic errors due to potential $O(\Lambda a)$ term,
where $\Lambda$ is some physical mass scale in the theory. These terms
are not necessarily small, even if the limit $a/L \to 0$ is
considered. Moreover, the lack of a perturbative expansion prevents us
defining the running coupling properly. The conclusion is that results
for $\bar g>\bar g^*$ could be affected by non-universal lattice
artefacts.

Studies of the running couplings in the SF scheme are a useful tool to
expose the possible existence of theories that show a conformal
behaviour at large distances. However, the results of numerical
simulations have to be interpreted with care; they are unlikely to
provide conclusive evidence about the existence of a fixed point by
themselves, but they can be used to check the consistency of scenarios
where the long--range dynamics is dictated by an IRFP. A more
convincing picture can emerge when these analyses are combined with
spectral
studies~\cite{DelDebbio:2008zf,Catterall:2008qk,Hietanen:2008mr,Pica:2009hc},
or MCRG methods~\cite{Hasenfratz:2009ea}.

\begin{figure}[H]
  \centering
  \epsfig{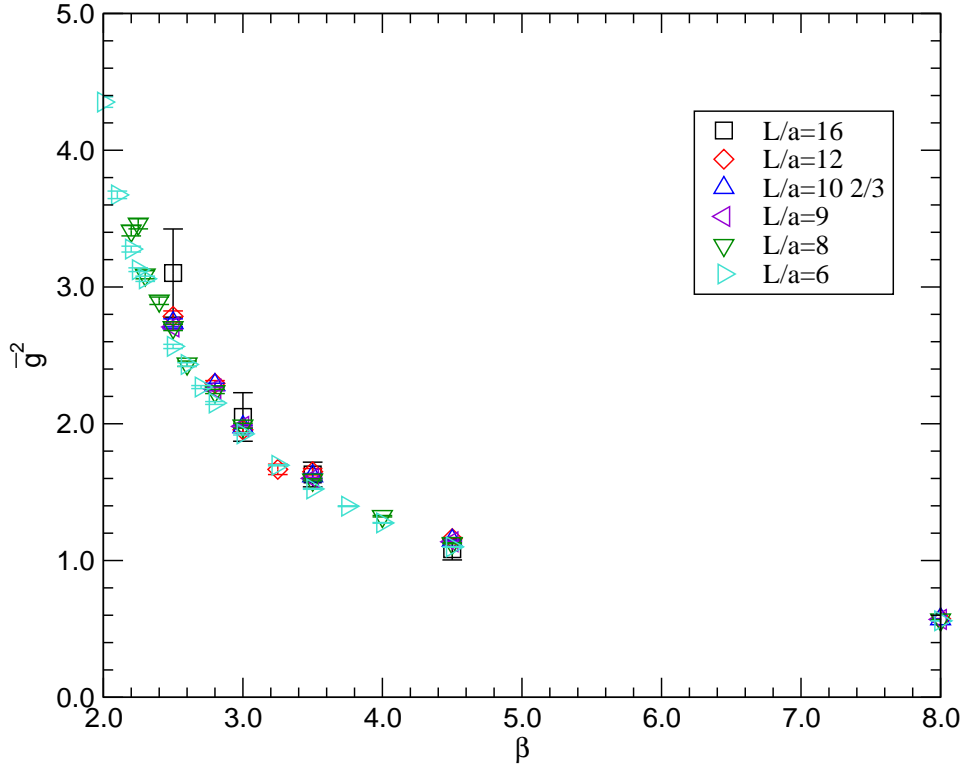}
  \caption{Data for the running coupling as computed from lattice
    simulations of the Schr\"odinger functional. Numerical simulations
    are performed at several values of the bare coupling $\beta$, and
    for several lattice resolutions $L/a$. The points at $L/a=9,10\frac{2}{3}$ are interpolated.
}
  \label{fig:SFdata}
\end{figure}

\sect{Results for the coupling}
\label{sec:SFcoupling}
We have measured the coupling $\overline{g}^2(\beta,L)$ for a range of
$\beta,L$. Our results are reported in Tab.~\ref{tab:gdata}, and
plotted in Fig.~\ref{fig:SFdata}: it is clear that the coupling is
very similar for different $L/a$ at a given value of $\beta$, and
hence that it runs slowly.

\begin{figure}[H]
  \centering 
  \epsfig{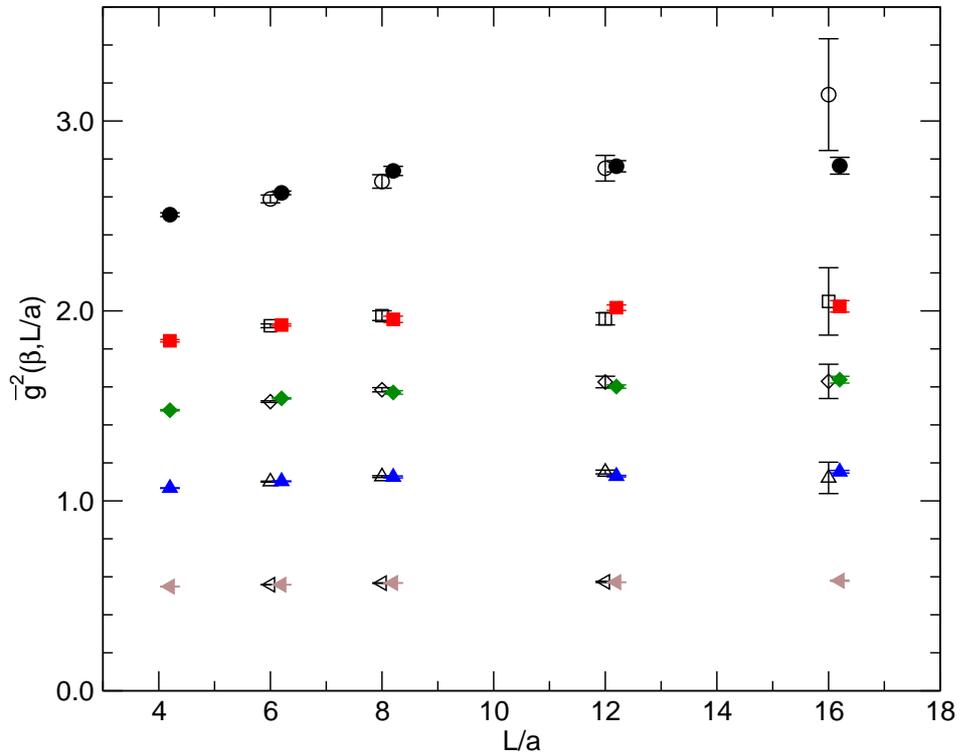}
  \caption{The results of our numerical simulations are compared to
    recent results obtained in Ref.~\cite{Hietanen:2009az}. Different
    symbols correspond to different values of the lattice bare coupling
    $\beta$, corresponding respectively to
    $\beta=2.5,3.0,3.5,4.5,8.0$. Empty symbols correspond to the data
    obtained in this work. Full symbols correspond to the data in
    Ref.~\cite{Hietanen:2009az}. Symbols have been shifted
    horizontally for easier reading of the plot. }
  \label{fig:karidata}
\end{figure}

In Fig.~\ref{fig:karidata} we compare our results to those obtained in
Ref.~\cite{Hietanen:2009az}. Our results are directly comparable since
we use the same action and definition of the running coupling, and it
is reassuring to see that they agree within statistical errors. The
numbers reported in the figure have been obtained using completely
independent codes; they constitute an important sanity check at these
early stages of simulating theories beyond QCD. 

The running of the coupling is encoded in the step scaling
function $\sigma(u,s)$ as
\begin{eqnarray}
  \label{eq:Sig}
  \Sigma(u,s,a/L)&=&\left. \overline{g}^2(g_0,sL/a)
  \right|_{\overline{g}^2(g_0,L/a)=u}\, , \\
  \label{eq:sigma}
  \sigma(u,s) &=& \lim_{a/L\to 0} \Sigma(u,s,a/L)\, ,
\end{eqnarray}
as described in Ref.~\cite{Luscher:1992an}.  The function
$\sigma(u,s)$ is the continuum extrapolation of $\Sigma(u,s,a/L)$
which is calculated at various a/L, according to the following
procedure. Actual simulations have been performed at the values of $\beta$
and $L$ reported in Tab.~\ref{tab:gpar}.

\begin{table}[H]
  \centering

{\scriptsize
  \begin{tabular}{|c|c|c|c|c|}
    \hline
    $\beta$ & $L$=6 & $L$=8 & $L$=12 & $L$=16\\
    \hline
    2.00 & 4.237(58) & - & - & - \\
    2.10 & 3.682(39) & - & - & - \\
    2.20 & 3.262(31) & 3.457(59) & - & - \\
    2.25 & 3.125(19) & 3.394(54) & - & - \\
    2.30 & 3.000(25) & 3.090(46) & - & - \\
    2.40 & 2.813(21) & 2.887(44) & - & - \\
    2.50 & 2.590(20) & 2.682(35) & 2.751(68) &  3.201(324) \\
    2.60 & 2.428(16) & 2.460(29) & - & - \\
    2.70 & 2.268(14) & - & - & - \\
    \hline
  \end{tabular}
  \begin{tabular}{|c|c|c|c|c|}
    \hline
    $\beta$ & $L$=6 & $L$=8 & $L$=12 & $L$=16\\
    \hline
    2.80 & 2.141(12) & 2.218(22) & 2.309(40) & - \\
    3.00 & 1.922(10) & 1.975(25) & 1.958(32) &  2.025(157) \\
    3.25 & 1.694(5) & - & 1.830(90) & - \\
    3.50 & 1.522(4) & 1.585(11) & 1.626(30) & 1.603(76) \\
    3.75 & 1.397(3) & - & - & - \\
    4.00 & 1.275(3) & 1.320(7) & - & - \\
    4.50 & 1.101(3) & 1.128(5) &  1.152(10) & 1.106(64) \\
    8.00 & 0.558(1) & 0.567(2) & 0.574(3) & - \\
    & & & & \\
    \hline
  \end{tabular}
}

\caption{Measured values of $\overline{g}^2$ on different volumes as a function of the bare coupling $\beta$.
}

\label{tab:gdata}
\end{table}

Starting from the actual data, we interpolate quadratically in $a/L$
to find values of $\overline{g}^2(\beta,L)$ at $L=9,10\frac{2}{3}$, so
that we obtain data for four steps of size $s=4/3$ for $L\rightarrow
sL$: $L=6,8,9,12$; $sL=8,10\frac{2}{3},12,16$. Then for each L we
perform an interpolation in $\beta$ using the same functional form as
Ref.~\cite{Appelquist:2009ty}:
\begin{equation}
\label{eq:fitg}
\frac{1}{\overline{g}^2(\beta,L/a)} = 
\frac{\beta}{2N}\left[\sum_{i=0}^{n} c_{i} \left(\frac{2N}{\beta}\right)^{i} \right]
\end{equation}
We choose to truncate the series with the number of parameters that
minimises the $\chi^2$ per degree of freedom. 

\begin{figure}[H]
  \centering
  \epsfig{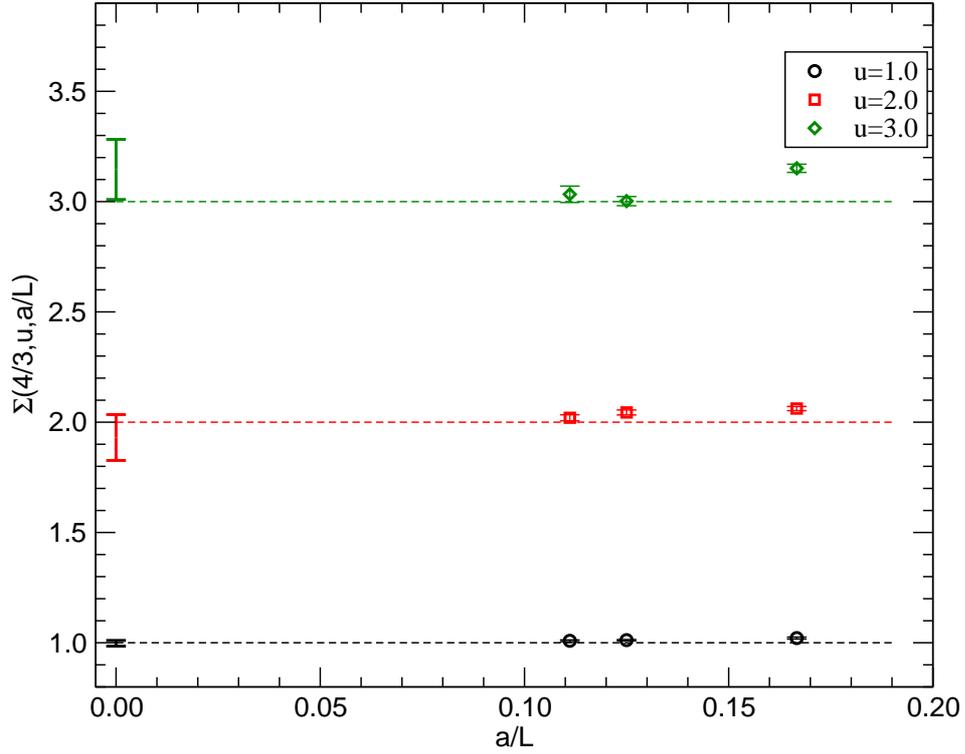}  
  \caption{Results for the lattice step--scaling function
    $\Sigma(4/3,u,a/L)$. The dashed lines represent the initial value
    of $u$. The point at $x=0$ yields the value of $\sigma(u)$, {\it
      i.e.} the extrapolation of $\Sigma$ to the continuum limit. The
    error bar shows the difference between constant and linear
    extrapolation functions, and gives an estimate of the systematic
    error in the extrapolation as discussed in the text.}
  \label{fig:SFSigma}
\end{figure}

All the subsequent analysis is based on these interpolating functions,
and does not make further use of the original data. Using the fitted
function in Eq.~(\ref{eq:fitg}), we compute $\Sigma(u,4/3,a/L)$ at a
number of points in the range $u \in [0.5,3.5]$.  A continuum
extrapolation is then performed in $a/L$ using these points to give a
single estimate of $\sigma(u)\equiv\sigma(u,4/3)$. Example
extrapolations for three values of $u$ are shown in
Fig.~\ref{fig:SFSigma}. The $L=6$ data were found to have large $O(a)$
artifacts, and are not used in the continuum extrapolation. The $L=16$
data have a large statistical error, which limits their current impact
on the continuum extrapolation. The sources of systematic uncertainty
in our final results for $\sigma(u)$ are due to the interpolation in
$L$ and $\beta$ and to the extrapolation to the continuum limit. Full
details of the statistical and systematic error analysis are provided
in Appendix~\ref{sec:app1}.

\begin{figure}[H]
  \centering
  \epsfig{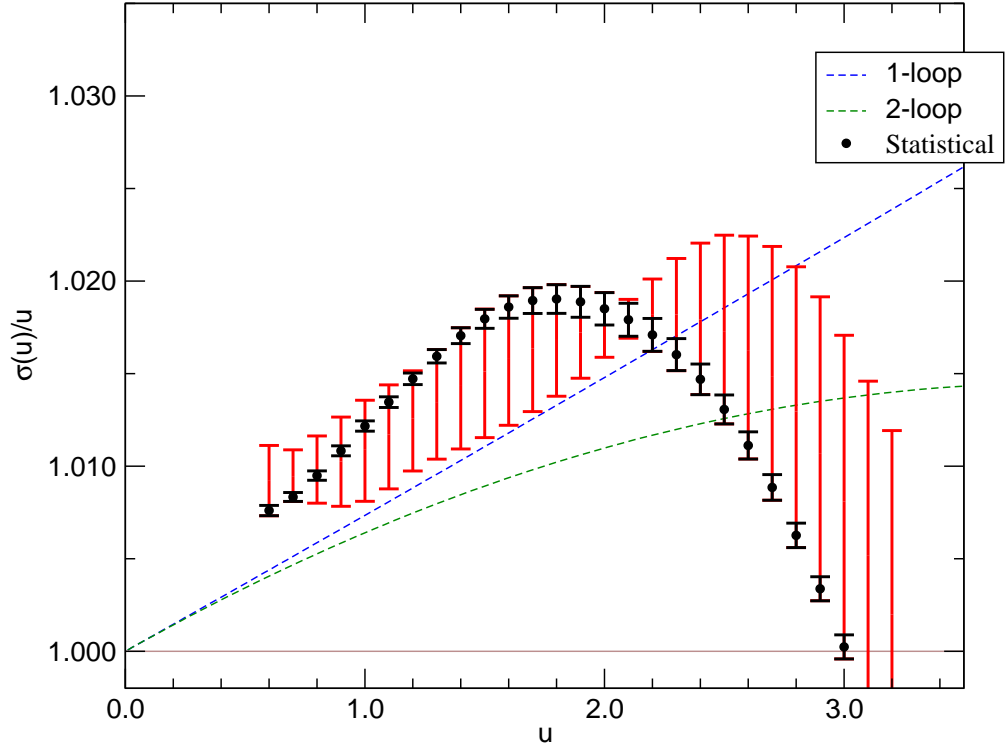}  
  \caption{The relative step--scaling function $\sigma(u)/u$ obtained
    after extrapolating the lattice data to the continuum limit. The black circles
have a statistical error only. The red error bars also include systematic
errors, but using only a constant continuum extrapolation (i.e. ignoring lattice artifacts). Note
    that a fixed point is identified by the condition $\sigma(u)/u=1$.
  }
  \label{fig:SFconstsigma}
\end{figure}

The resulting values for $\sigma(u)$ with statistical errors only can
be seen as the black circles in Fig.~\ref{fig:SFconstsigma}. The red
error bars in Fig.~\ref{fig:SFconstsigma} also include systematic
errors, but using only a constant continuum extrapolation. This is
equivalent to the assumption that lattice artefacts are negligible in
our data. A similar assumption has been used in
Ref.~\cite{Hietanen:2009az}, where the data at finite $a/L$ were used
directly to constrain the parameters that appear in the $\beta$
function of the theory. The study of the lattice step scaling
function, and its continuum extrapolation, that we employ for this
work, will ultimately allow us to obtain a full control over the
systematic errors.  

The step scaling function encodes the same information as the $\beta$
function. The relation between the two functions for a generic
rescaling of lengths by a factor $s$ is given by:
\begin{equation}
  \label{eq:stepbeta}
  -2 \log s = \int_{u}^{\sigma(u,s)} \frac{dx}{\sqrt{x}
    \beta(\sqrt{x})}\, .
\end{equation}
The step scaling function can be computed at a given order in
perturbation theory by using the analytic expression for the
perturbative $\beta$ function, and solving Eq.~(\ref{eq:stepbeta}) for
$\sigma(u,s)$.  On the other hand, it can be seen directly from the
definition of $\sigma(u,s)$ in Eq.~(\ref{eq:sigma}) that an IRFP
corresponds to $\sigma(u,s)=u$.

Our current values for the step scaling function are consistent with a
fixed point in the region $\overline{g}^2\sim2.0-3.2$, as reported in
Ref.~\cite{Hietanen:2009az}. Further simulation at higher
$\overline{g}^2$ is limited by the bulk transition observed in
Ref.~\cite{Catterall:2008qk,Hietanen:2008mr} at $\beta\simeq2.0$.

\begin{figure}[H]
  \centering
  \epsfig{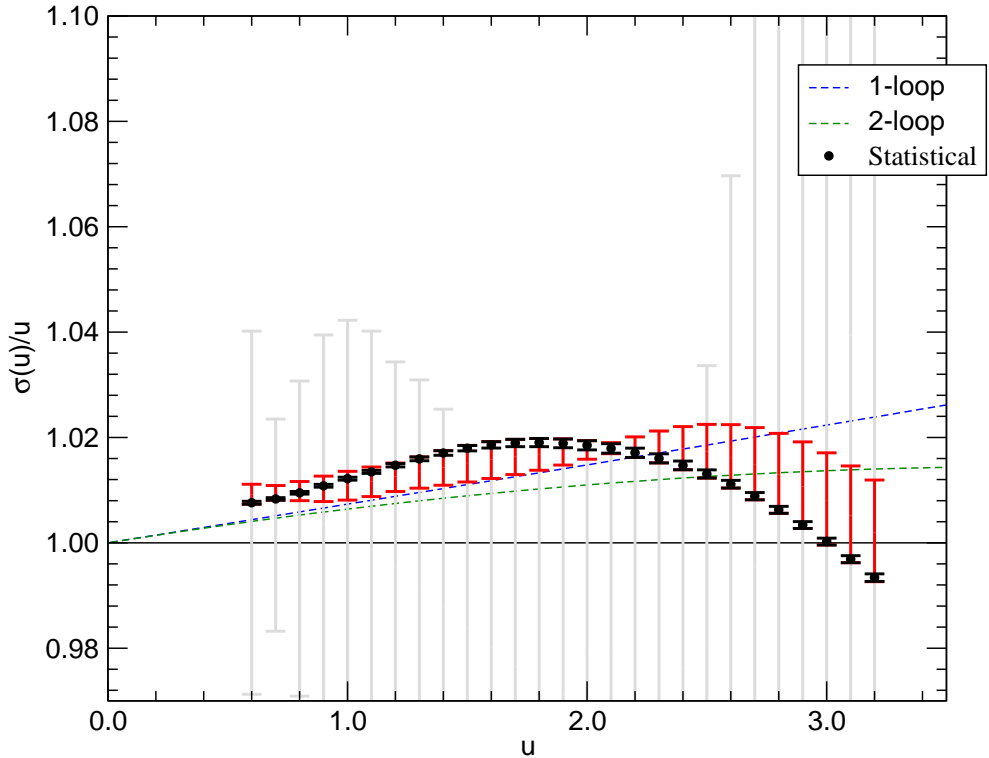}  
  \caption{The relative step--scaling function $\sigma(u)/u$ obtained
    after extrapolating the lattice data to the continuum limit. The
    black circles have a statistical error only, the red error bars include systematic errors but using only a constant continuum extrapolation, and the grey error bars give an idea of the total error by
    including both constant and linear continuum extrapolations.  }
  \label{fig:SFlinsigma}
\end{figure}

The errors from also including the linear continuum extrapolation are
much larger and mask any evidence for a fixed point, as shown in
Fig.~\ref{fig:SFlinsigma}. This should be a conservative estimate of
the total uncertainty on $\sigma(u)$, which is dominated by systematic
errors.

\sect{Running mass}
\label{sec:SFmass}
The running of the fermion mass is determined by the scale--dependence
of the renormalisation constant for the pseudoscalar fermion bilinear
$Z_P$ defined in Eq.~(\ref{eq:ZPdef}). Note that $Z_P$ is both scheme
and scale dependent.  The same step scaling technique described for
the gauge coupling can be used to follow the nonperturbative evolution
of the fermion mass in the SF scheme. In this work, we follow closely
the procedure outlined in Ref.~\cite{DellaMorte:2005kg}.

\begin{figure}[H]
  \centering
  \epsfig{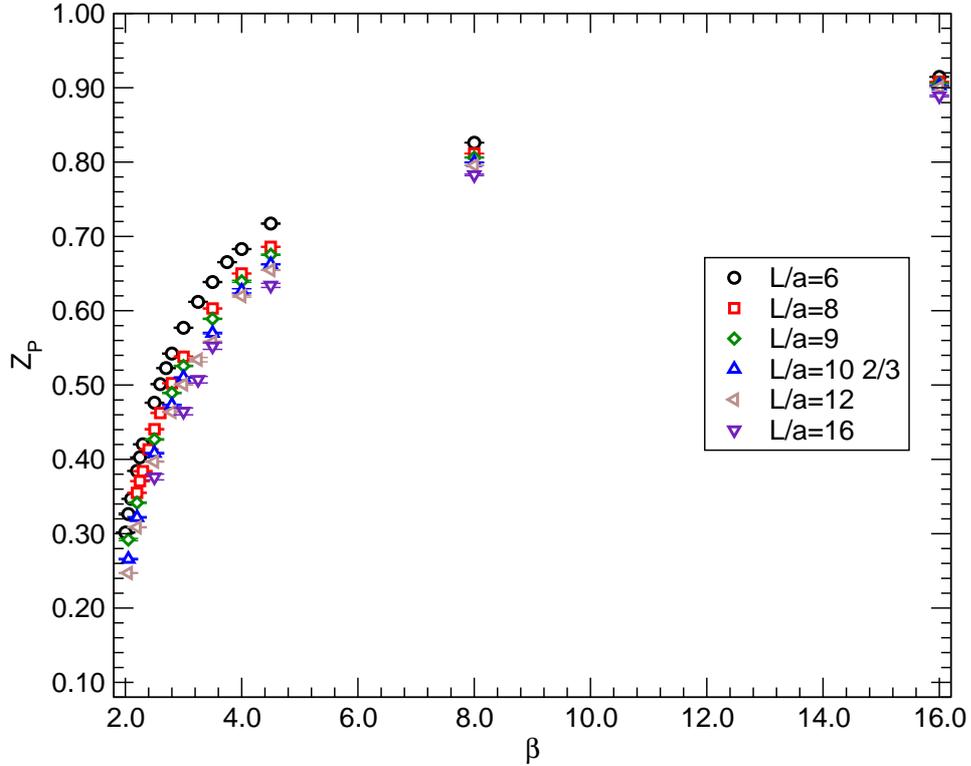}  
  \caption{Data for the renormalisation constant $Z_P$ as computed
    from lattice simulations of the Schr\"odinger
    functional. Numerical simulations are performed at several values
    of the bare coupling $\beta$, and for several lattice resolutions
    $L/a$. The points at $L/a=9,10\frac{2}{3}$ are interpolated.}
  \label{fig:ZPdata}
\end{figure}

We have measured the pseudoscalar density renormalisation constant
$Z_P(\beta,L)$ for a range of $\beta,L$.  Our results are reported in Tab.~\ref{tab:Zdata}, and plotted in
Fig.~\ref{fig:ZPdata}, where we see that there is a clear trend in
$Z_P$ as a function of $L$ at all values of $\beta$.

The lattice step scaling function for the mass is defined as:
\begin{equation}
  \label{eq:sigmaPdef}
  \Sigma_P(u,s,a/L)=\left
    . {\frac{Z_P(g_0,sL/a)}{Z_P(g_0,L/a)}}
  \right|_{\overline{g}^2(L)=u}\, ; 
\end{equation}
the mass step scaling function in the continuum limit,
$\sigma_P(u,s)$, is given by:
\begin{equation}
  \label{eq:sigma_p}
  \sigma_P(u,s) = \lim_{a\to 0}\Sigma_P(u,s,a/L)\, .
\end{equation}

\begin{table}[H]
  \centering

 {\scriptsize
  \begin{tabular}{|c|c|c|c|c|}
    \hline
    $\beta$ & $L$=6 & $L$=8 & $L$=12 & $L$=16\\
    \hline
    2.00 & 0.3016(6) & - & - & - \\
    2.05 & 0.3265(11) & - & 0.2466(6) & - \\
    2.10 & 0.3469(6) & - & - & - \\
    2.20 & 0.3845(6) & 0.3550(7) & 0.3087(6) & - \\
    2.25 & 0.4028(6) & 0.3707(7) & - & - \\
    2.30 & 0.4203(6) & 0.3841(7) & - & - \\
    2.40 & - & 0.4134(7) & - & - \\
    2.50 & 0.4762(6) & 0.4406(9) & 0.3970(7) & 0.3763(39) \\
    2.60 & 0.5012(7) & 0.4624(7) & - & - \\
    2.70 & 0.5228(6) & - & - & - \\
    \hline
  \end{tabular}
  \begin{tabular}{|c|c|c|c|c|}
    \hline
    $\beta$ & $L$=6 & $L$=8 & $L$=12 & $L$=16\\
    \hline
    2.80 & 0.5424(7) & 0.5025(6) & 0.4639(6) & - \\
    3.00 & 0.5770(7) & 0.5381(7) & 0.5008(8) & 0.4647(55) \\
    3.25 & 0.6120(6) & - & 0.5342(30) & 0.5063(44) \\
    3.50 & 0.6385(7) & 0.6030(7) & 0.5580(10) & 0.5523(43) \\
    3.75 & 0.6654(6) & - & - & - \\
    4.00 & 0.6830(6) & 0.6501(6) & 0.6197(14) & - \\
    4.50 & 0.7173(7) & 0.6859(6) & 0.6547(4) & 0.6341(27) \\
    8.00 & 0.8261(3) & 0.8114(3) & 0.7956(2) & 0.7827(11) \\
    16.0 & 0.9146(4) & 0.9082(2) & 0.9005(5) & 0.8887(15) \\
     & & & &  \\
    \hline
  \end{tabular}
}

\caption{Measured values of $Z_P$ on different volumes as a function of the bare coupling $\beta$.
}
\label{tab:Zdata}
\end{table}

The method for calculating $\sigma_P(u)\equiv\sigma_P(u,4/3)$ is
similar to that outlined in Sec.~\ref{sec:SFcoupling} for calculating
$\sigma(u)$. Interpolation in $\beta$ is accomplished using a
function of the form:
\begin{equation}
\label{eq:fitZ_p}
Z_{P}(\beta, L/a) = \sum_{i=0}^{n} c_{i}\left(\frac{1}{\beta}\right)^{i}
\end{equation}

\begin{figure}[H]
  \centering
  \epsfig{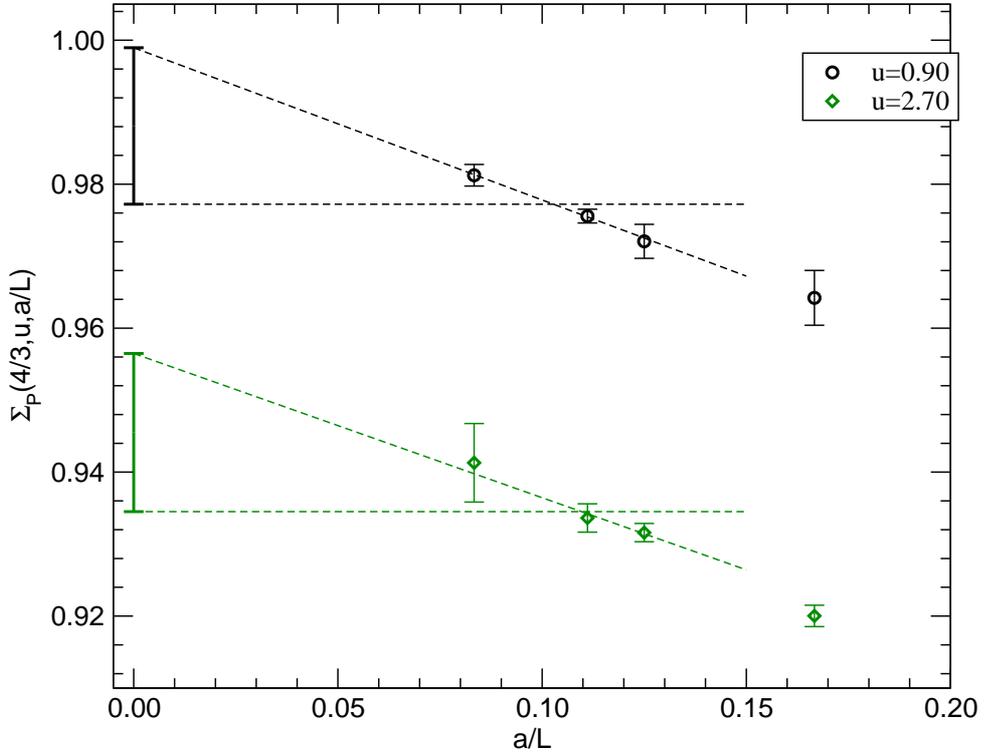}  
  \caption{Results for the lattice step--scaling function
    $\Sigma_P(4/3,u,a/L)$. The point at $x=0$ yields the value of
    $\sigma_P(u)$, {\it i.e.} the extrapolation of $\Sigma_P$ to the
    continuum limit. The error bar shows the difference between
    constant and linear extrapolation functions, and gives an estimate
    of the systematic error in the extrapolation as discussed in the
    text.}
  \label{fig:SFsigmaP_extrap}
\end{figure}

Full details of the procedure are given in
Appendix~\ref{sec:app2}. Again the errors are dominated by
systematics, in particular the choice of continuum extrapolation
function. In Fig.~\ref{fig:SFsigmaP_extrap} we see that, unlike
$\overline{g}^2$, $Z_P$ has a significant variation with $a/L$ that is
fit well by a linear continuum extrapolation. The constant
extrapolation is only used to quantify the errors in extrapolation.

\begin{figure}[H]
  \centering
  \epsfig{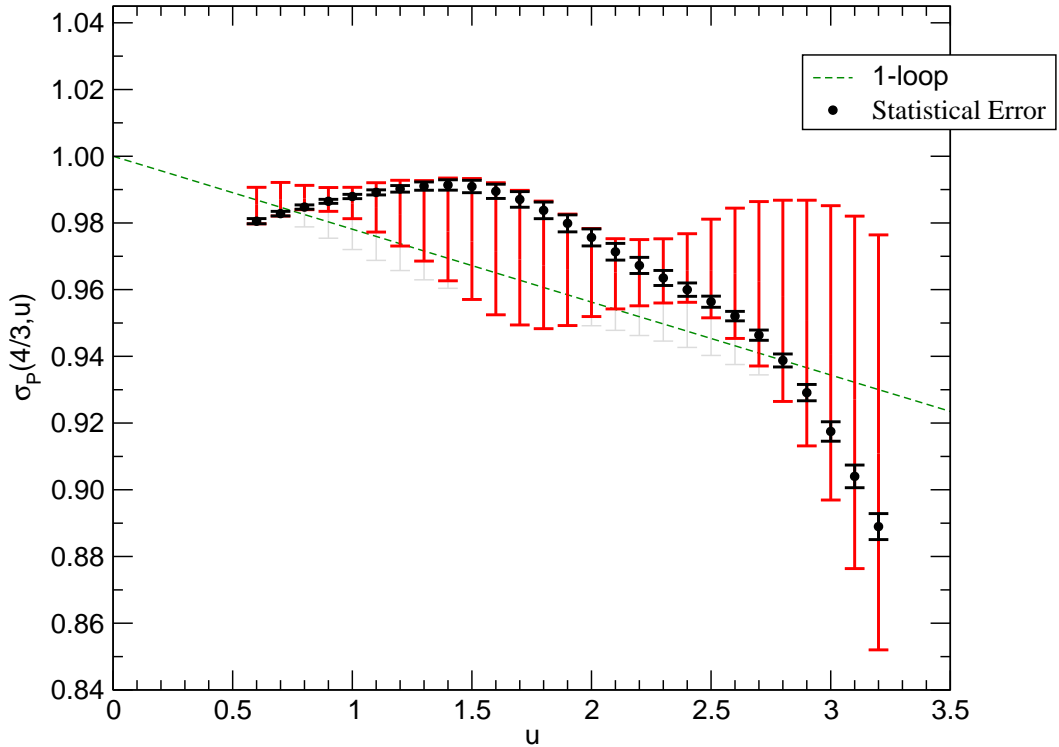}  
  \caption{The step-scaling function for the running mass
    $\sigma_P(u)$, using a linear continuum extrapolation. The
    black circles have a statistical error only, the red error bars include systematic errors using a
    linear continuum extrapolation. The grey
    error bars come from also including a constant extrapolation of
    the two points closest to the continuum, and give an idea of the
    systematic error in the continuum extrapolation.}
  \label{fig:sigmaP}
\end{figure}

Using the fact that $\sigma_P(u,s) =
\overline{m}(\mu)/\overline{m}(\mu/s)$ for $\mu=1/L$, we can perform
an iterative step scaling of the coupling and the mass to determine
the running of the mass with scale. However, since we observe no
running of the coupling within errors this is not particularly
interesting.

The mass step scaling function is related to the mass anomalous
dimension (see e.g. Ref.~\cite{DellaMorte:2005kg}):
\begin{equation}
  \label{eq:sigPgamma}
  \sigma_P(u) = \left(\frac{u}{\sigma(u)}\right)^{(d_0/(2\beta_0))}
  \exp\left[\int_{\sqrt{u}}^{\sqrt{\sigma(u)}} dx 
    \left(\frac{\gamma(x)}{\beta(x)}-\frac{d_0}{\beta_0 x}\right)\right]\, .
\end{equation}
We find good agreement with the 1-loop perturbative prediction, as
shown in Fig.~\ref{fig:sigmaP}.

In the vicinity of an IRFP the relation between $\sigma_P$ and
$\gamma$ simplifies. Denoting by $\gamma^*$ the value of the anomalous
dimension at the IRFP, we obtain:
\begin{equation}
  \label{eq:gammafix}
  \int_{\overline{m}(\mu)}^{\overline{m}(\mu/s)} \frac{dm}{m} = 
  -\gamma^* \int_{\mu}^{\mu/s} \frac{dq}{q}\, ,
\end{equation}
and hence:
\begin{equation}
  \label{eq:sigmaPfix}
  \log\left|\sigma_P(s,u)\right| = -\gamma^* \log s\, . 
\end{equation}
We can therefore define an estimator
\begin{equation}
\label{eq:tau}
\hat\gamma(u) =
-\frac{\log\left|\sigma_P(u,s)\right|}{\log\left|s\right|}\, ,
\end{equation}
which yields the value of the anomalous dimension at the fixed
point. Away from the fixed point $\hat\gamma$ will deviate from the
anomalous dimension, with the discrepancy becoming larger as the
anomalous dimension develops a sizeable dependence on the energy
scale. 

\begin{figure}[H]
  \centering
  \epsfig{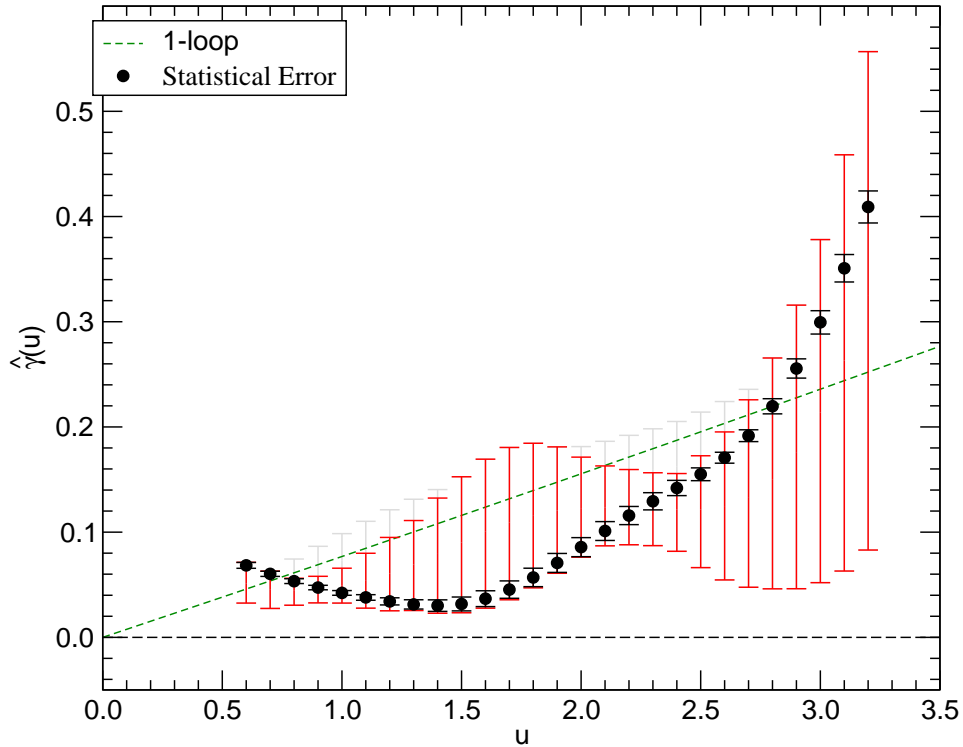}  
  \caption{The mass anomalous dimension estimator $\hat\gamma(u)$. The dashed line
    shows the 1-loop perturbative result, the
    black circles have a statistical error only, and the red error bars include systematic errors using a
    linear continuum extrapolation. The grey error
    bars also include a constant extrapolation of
    the two points closest to the continuum, giving an idea
    of the systematic error involved in the continuum extrapolation.}
  \label{fig:gamma}
\end{figure}

We plot the estimator $\hat\gamma$ in Fig.~\ref{fig:gamma}. Again the
error bars come from evaluating the above expression using the
extremal values of $\sigma_P(u)$ at each $u$.  We see that the actual
value of $\hat\gamma$ is rather small over the range of interest.  In
particular at $\overline{g}^2=2.2$, the benchmark value for the IRFP
tentatively found in Ref.~\cite{Hietanen:2009az}, we have
$\hat\gamma=0.116^{+43}_{-28}$ using just the linear continuum
extrapolation, and $\hat\gamma=0.116^{+76}_{-28}$ if we include the
constant continuum extrapolation as well. In the presence of an IRFP
$\hat\gamma$ yields the value of the anomalous dimension, and
therefore the values above can be used to bound the possible values of
$\gamma^*$. The results of Ref.~\cite{Hietanen:2009az} suggest the
IRFP is in the range $\overline{g}^2=2.0-3.2$; at the extremes of this
range we find $\gamma^*=0.086^{+85}_{-10}$ and $0.41^{+15}_{-33}$ using
just the linear continuum extrapolation, and
$\gamma^*=0.086^{+105}_{-10}$ and $0.41^{+15}_{-33}$ including the
constant continuum extrapolation. Over the entire range of
couplings consistent with an IRFP, $\gamma^*$ is constrained to lie in
the range $0.05 < \gamma^* < 0.56$, even with our more conservative
assessment of the continuum extrapolation errors.

\sect{Conclusions}
\label{sec:concl}
In this paper we have presented results for the running of the
Schr\"{o}dinger Functional coupling $\overline{g}^2$ and the mass
anomalous dimension $\gamma$.

Turning first to the running of the coupling, our results are
completely consistent with those of Ref.~\cite{Hietanen:2009az}. Our
statistical errors are larger; however, we have carried out our
analysis in a way that aims at disentangling clearly the scale
dependence from the lattice artefacts. Our analysis can be
systematically improved as more extensive studies are performed, and
will ultimately allow us to take the continuum limit with full control
over the resulting systematic errors. Our results appear to show a
slowing in the running of the coupling above $\overline{g}^2=2$ or so,
and are consistent with the presence of a fixed point where the
running stops at somewhat higher $\overline{g}^2$. This is consistent
with the analysis of Ref.~\cite{Hietanen:2009az}. However, once we
include the systematic errors from the continuum extrapolation we find
that our results no longer give any evidence for a fixed point. The
fundamental reason for this is that the running of the coupling is
very slow in this theory and so great accuracy is needed, in
particular near a possible fixed point.

By contrast, we find that the behaviour of the anomalous dimension
$\gamma$ is much easier to establish. The systematic errors from the
continuum extrapolation are much smaller than the signal, and we find
a moderate anomalous dimension, close to the 1-loop perturbative
prediction, throughout the range of $\beta$ explored. In particular,
in the range $\overline{g}^2=2.0-3.2$, where there may be an infrared
fixed point, we find $0.05 < \gamma < 0.56$. These values are much
smaller than those required for phenomenology, which are typically of
order 1-2. Such large values of $\gamma$ are clearly inconsistent with
our results. The anomalous dimension at the fixed point can be
computed analytically using the all--order beta function proposed in
Ref.~\cite{Ryttov:2007cx}. The result can be expressed as a function
of group--theoretical factors only. Using the conventions described in
the Appendix of Ref.~\cite{DelDebbio:2008wb} for these
group--theoretical factors, the result in Ref.~\cite{Ryttov:2007cx}
yields $\gamma=3/4$, which is not too far from the bound we quote
above. Given the uncertainty in the exact value of the SF coupling at
the fixed point, $\gamma=3/4$ is not strongly excluded by our current
data. More precise investigations are needed to clarify this
point. 

The anomalous dimension is easier to measure than the beta function in
candidate walking technicolor theories since it is expected to be
different from zero, so we are measuring the difference of two
quantities that are significantly different, say $Z_P$ at $L=8$ and
$L=12$. By contrast for the running of the coupling we must measure
the difference of two quantities that are almost the same, since the
beta function is expected to be small. Furthermore, the anomalous
dimension is crucial for phenomenology; if it is not large then the
presence or absence of walking behaviour becomes academic. Hence the
implications of our measurement of $\gamma$ for the phenomenology of
minimal walking technicolor call for a more precise study. 

Our conclusion that $\gamma$ is not large is unlikely to be affected
by using larger lattices. One can see this by considering the
continuum extrapolations in Fig.~\ref{fig:SFsigmaP_extrap}. For
$\gamma$ to reach, say, 1 in the continuum limit, we would need
$\Sigma_P$ to be $3/4=0.75$ at $a/L=0$. However we see that the
dependence on $a/L$ is much too small for this to be possible, and
indeed is in the wrong direction. Only a very unlikely conspiracy of
lattice artifacts would make it possible for $\Sigma_P$ to be as small
as 0.75 in the continuum limit. On the other hand the value of $\bar
g$ corresponding to the IRFP is currently not known with
sufficient accuracy.

The results presented here are the first computation of the anomalous
dimension at a putative fixed point; the systematic errors need to be
reduced to make our conclusions more robust. In particular, using
larger lattices would give results at smaller $a/L$ and hence make the
continuum extrapolations more accurate. It may also be necessary to
use an improved action in the long term to achieve the precision
required to show the existence of an IRFP or of walking behaviour.
However, as described above, this is very unlikely to affect our
phenomenologically most important result, namely that $\gamma$ is not
large. Recent results in Ref.~\cite{DeGrand:2009hu} suggest that the
anomalous dimension can be computed using finite--size scaling
techniques. A comparison of different techniques will improve the
determination of the anomalous dimension.

\section*{Acknowledgments}
\label{sec:ack}

We thank Ari Hietanen, Kari Rummukainen, and Kimmo Tuominen for useful
discussions, and for providing access to their data. We also thank
Francesco Sannino for discussions on the anomalous dimension at an
IRFP. This work has made use of the Darwin Supercomputer of the
University of Cambridge High Performance Computing Service
(http://www.hpc.cam.ac.uk/), provided by Dell Inc. using Strategic
Research Infrastructure Funding from the Higher Education Funding
Council for England; resources funded by the University of Oxford and
EPSRC; the PC cluster at the University of Southern Denmark; resources
provided by the Edinburgh Compute and Data Facility (ECDF)
(http://www.ecdf.ed.ac.uk/). The ECDF is partially supported by the
eDIKT initiative ( http://www.edikt.org.uk). LDD is supported by an
STFC Advanced Fellowship.

\appendix

\sect{Coupling error analysis}
\label{sec:app1}
We directly measure the Schr\"odinger Functional coupling
$\overline{g}^2$ and perform multiple stages of interpolation and
extrapolation to extract the continuum step scaling function $\sigma(u)\equiv\sigma(u,4/3)$.

\begin{table}[H]
  \centering

\begin{scriptsize}
\[
\begin{array}{|c||c|c|c|c|c|}
\hline
\overline{g}^2 & \multicolumn{5}{|c|}{L/a} \\
\hline
 & 6 & 8 & 9 & 10\frac{2}{3} & 12\\
\hline
c_0 & 1.113\pm0.057 & 0.967\pm0.050 & 1.010\pm0.001 & 0.987\pm0.003 & 0.988\pm0.024 \\
c_1 & -0.560\pm0.206 & -0.064\pm0.215 & -0.259\pm0.001 & -0.216\pm0.006 & -0.226\pm0.055 \\
c_2 & 0.130\pm0.216 & -0.307\pm0.328 &  & -0.022\pm0.003 & -0.016\pm0.028 \\
c_3 & 0.366\pm0.125 & 0.221\pm0.211 &  &  &  \\
c_4 & -0.136\pm0.196 & -0.059\pm0.048 &  &  &  \\
c_5 & -0.364\pm0.234&  &  &  &  \\
c_6 & 0.298\pm0.127 &  &  &  &  \\
c_7 & -0.064\pm0.024 &  &  &  &  \\
\hline
\frac{\chi^2}{dof} & 2.85 & 2.42 & 1.73 & 3.45 & 3.37 \\
dof & 8 & 7 & 4 & 3 & 4 \\
\hline
\end{array}
\]
\end{scriptsize}

\caption{Interpolation best fit parameters for $\overline{g}^2$.
}
\label{tab:gfit1}
\end{table}

\begin{figure}[H]
  \centering
  \epsfig{file=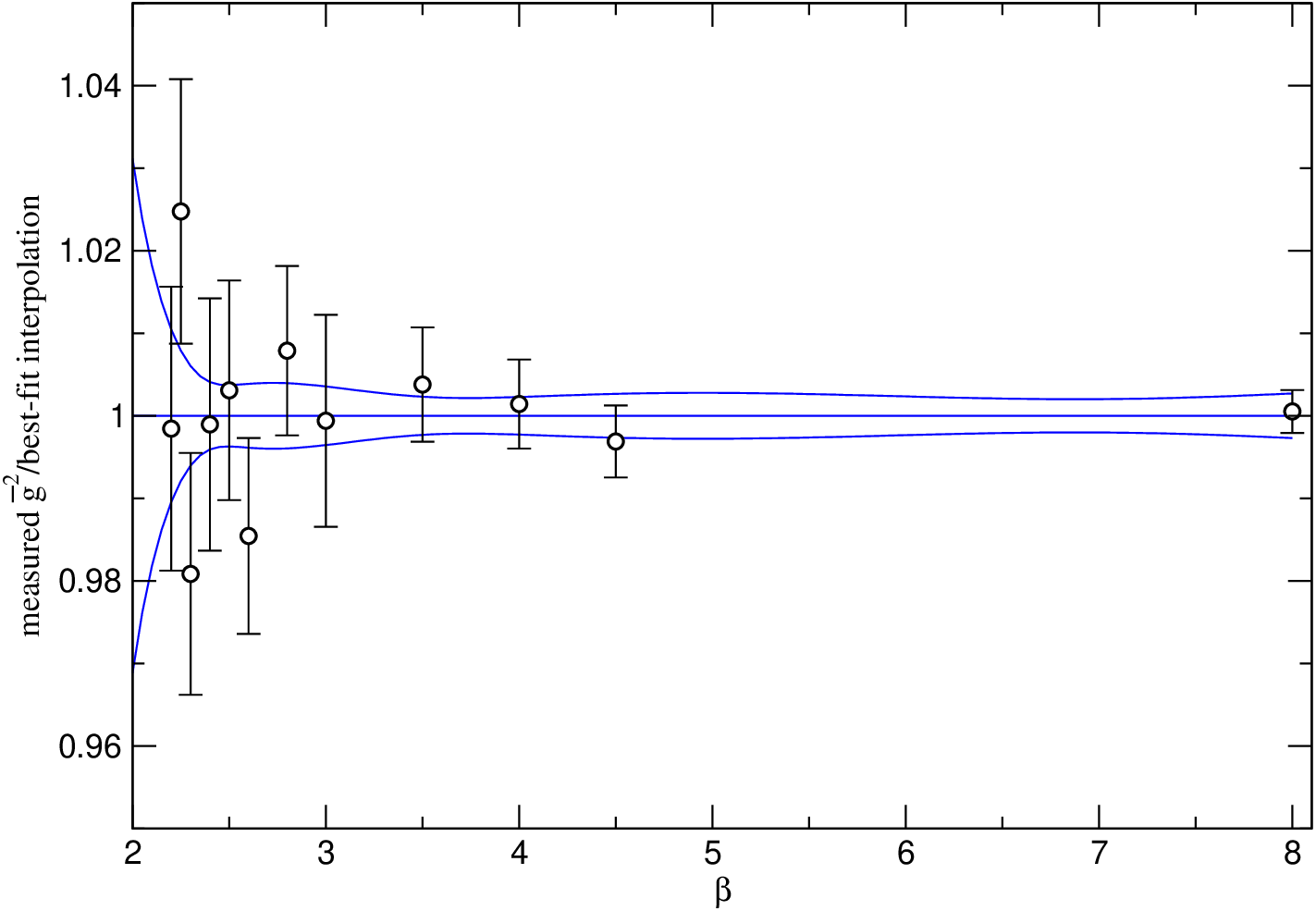,scale=0.6,clip}  
  \caption{Example of an interpolation function for $L=8$, with a $\pm\sigma$ confidence interval, compared with measured $\overline{g}^2$ data points.}
  \label{fig:SFg_fit}
\end{figure}

In order to estimate our errors for each of these stages we perform
multiple bootstraps of the data. The full procedure to get a single
estimate of $\sigma(u)$ can be summarised as follows:

\begin{itemize}
\item Generate $N_b \times N_a$ bootstrapped ensembles of the data and
  extract mean and error for each.
\item For each bootstrap, interpolate in $a/L$ to find values at
  $L=9,10\frac{2}{3}$.
\item From each set of $N_a$ of these find the mean and standard
  deviation, to give $N_b$ interpolated data points with error bars.
\item For each of the $N_b$ bootstraps do a non-linear least squares
  fit for $\overline{g}^2(\beta,L)$ interpolation functions in
  $\beta$, an example is shown in Fig.~\ref{fig:SFg_fit}.
\item Use these functions to find $N_b$ estimates of $\Sigma(a/L,u)$
  for $L=8,9$, and from this extract a mean and error for each $a/L$.
\item Perform a single weighted continuum extrapolation in $a/L$ using
  these points to give $\sigma(u)$.
\end{itemize}

This process is repeated $N_m$ times, bringing the total number of
bootstrap replicas of the data to $N_a \times N_b \times N_m$. This
gives $N_m$ estimates of $\sigma(u)$, from which a mean and 1-sigma
confidence interval is extracted.

However, the systematic errors that
result from varying the number of parameters in the interpolation
functions or the continuum extrapolation functions are significantly
larger than the statistical errors for the optimal set of parameters.

In order to quantify this, we repeated the entire bootstrapped process
of calculating $\sigma(u)$ with a range of different interpolation and
extrapolation functions, each of which gives an estimates for
$\sigma(u)$, with a statistical error. 

Specifically, we included two different choices for the number of
parameters in the interpolating functions at each $L$. We kept the
best fit, outlined in Tab.~\ref{tab:gfit1} and added the function with
the second lowest $\chi^2$ per degree of freedom as shown in
Tab.~\ref{tab:gfit2}. The error in the continuum extrapolation was
estimated by including both constant and linear extrapolation
functions.  All possible combinations of these functions gave us a set
of $2^{5}=32$ values for $\sigma(u)$, each with a statistical error,
which spanned the range of the systematic variation.

For each value of $u$ the resulting extremal values of $\sigma(u)$
were used as upper and lower bounds on the central value. 

\begin{table}[H]
  \centering
  \begin{scriptsize}
    \[
    \begin{array}{|c||c|c|c|c|c|}
      \hline
      \overline{g}^2 & \multicolumn{5}{|c|}{L/a} \\
      \hline
      & 6 & 8 & 9 & 10\frac{2}{3} & 12\\
\hline
c_0 & 1.113\pm0.057 & 0.967\pm0.050 & 1.010\pm0.001 & 0.987\pm0.003 & 0.988\pm0.024 \\
c_1 & -0.560\pm0.206 & -0.064\pm0.215 & -0.259\pm0.001 & -0.216\pm0.006 & -0.226\pm0.055 \\
c_2 & 0.130\pm0.216 & -0.307\pm0.328 &  & -0.022\pm0.003 & -0.016\pm0.028 \\
c_3 & 0.366\pm0.125 & 0.221\pm0.211 &  &  &  \\
c_4 & -0.136\pm0.196 & -0.059\pm0.048 &  &  &  \\
c_5 & -0.364\pm0.234&  &  &  &  \\
c_6 & 0.298\pm0.127 &  &  &  &  \\
c_7 & -0.064\pm0.024 &  &  &  &  \\
\hline
      \frac{\chi^2}{dof} & 2.85 & 2.42 & 1.73 & 3.45 & 3.37 \\
      dof & 8 & 7 & 4 & 3 & 4 \\
      \hline
    \end{array}
    \]
  \end{scriptsize}

  \caption{Interpolation next-best fit parameters for $\overline{g}^2$.
  }
  \label{tab:gfit2}
\end{table}

\begin{figure}[H]
  \centering
  \epsfig{file=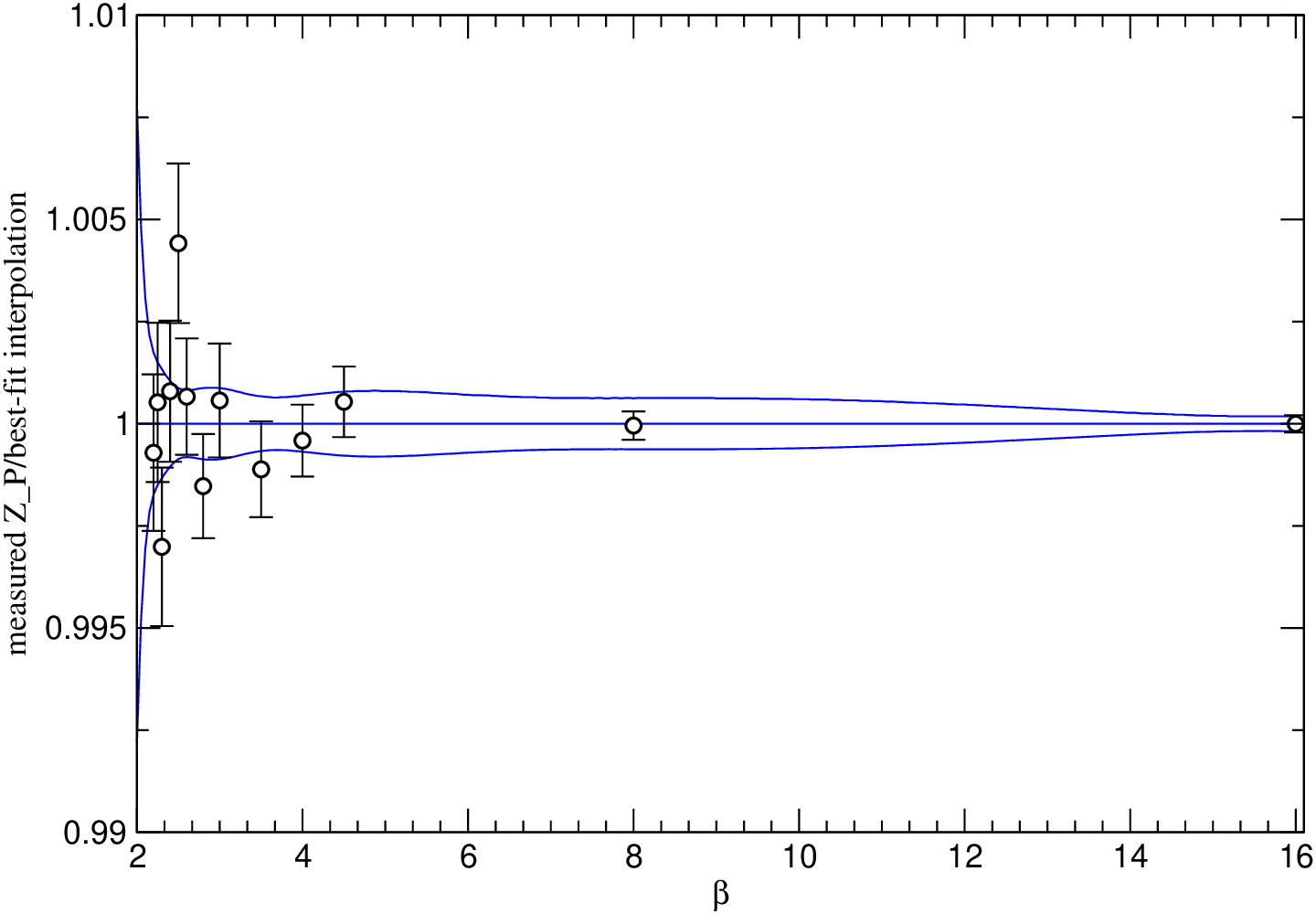,scale=0.6,clip}  
  \caption{Example of an interpolation function for $L=8$, with a $\pm\sigma$ confidence interval, compared with measured $Z_P$ data points.}
  \label{fig:SFZ_fit}
\end{figure}

\sect{Mass error analysis}
\label{sec:app2}
The mass error analysis follows the same procedure as outlined in
Appendix~\ref{sec:app1} with $\overline{g}^2$ replaced by $Z_P$. The function used to interpolate $Z_P$
in $\beta$ is given in Eq.~\ref{eq:fitZ_p}, and an example fit is shown in Fig.~\ref{fig:SFZ_fit}.
The $c_i$ giving the smallest reduced $\chi^2$ are given in
Tab.~\ref{tab:zfit1} and those with the second smallest in
Tab.~\ref{tab:zfit2}. 

 In addition, $Z_P$ converges faster than
$\overline{g}^2$ and we have better $16^4$ data so we can use 3 points
in our continuum extrapolations. Again the $L=6$
data were found to have large $O(a)$ artifacts so are not used in the continuum extrapolation, and for the constant extrapolation only the two points closest to the continuum limit are used.
The fits for both $\overline{g}^2$
and $Z_P$ are required to determine $\sigma_P(u)$, so independently
varying the choice of the number of parameters for these now gives
$2^{10}=1024$ values for $\sigma_P(u)$, each with a statistical error.

\begin{table}[H]
  \centering
\begin{scriptsize}
\[
\begin{array}{|c||c|c|c|c|c|c|}
\hline
Z_P & \multicolumn{6}{|c|}{L/a} \\
\hline
 & 6 & 8 & 9 & 10\frac{2}{3} & 12 & 16\\
\hline
c_0 &	0.58	\pm	0.30	&0.93	\pm	0.09	&1.02	\pm	0.01	&1.00	\pm	0.01	&1.01	\pm	0.01	&1.01	\pm 0.01 \\
c_1 &	7.64	\pm	6.85	&-0.43	\pm	1.74	&-2.17	\pm	0.10	&-1.76	\pm	0.01	&-1.98	\pm	0.08	&-1.99	\pm	0.09 \\
c_2 &	-78.87	\pm	60.50	&-8.18	\pm	12.64	&4.70	\pm	0.54	&1.56	\pm	0.05	&2.30	\pm	0.31	&1.93	\pm	0.43	\\
c_3 &	361.79	\pm	272.14	&36.42	\pm	43.33	&-10.73	\pm	1.27	&-2.14	\pm	0.06	&-3.01	\pm	0.34	&-2.23	\pm	0.64	\\
c_4 &	-898.23	\pm	662.83	&-75.69	\pm	71.04	&7.96	\pm	1.06	&	 & 	& \\		
c_5 &	1137.79	\pm	833.32	&57.07	\pm	44.83	&  &	  &	 &\\				
c_6 &	-579.79	\pm	424.25	&	  &	  &	  &	& \\						
\hline
\frac{\chi^2}{dof} & 2.42 & 1.66 & 2.24 & 4.82 & 6.68 & 6.67\\
dof & 11 & 8 & 5 & 6 & 6 & 3\\
\hline
\end{array}
\]
\end{scriptsize}

\caption{Interpolation best fit parameters for $Z_P$.
}
\label{tab:zfit1}
\end{table}

\begin{table}[H]
  \centering
\begin{scriptsize}
\[
\begin{array}{|c||c|c|c|c|c|c|}
\hline
Z_P & \multicolumn{6}{|c|}{L/a} \\
\hline
 & 6 & 8 & 9 & 10\frac{2}{3} & 12 & 16\\
\hline
c_0 & 1.00 \pm 0.07 & 1.14 \pm 0.46 & 0.89 \pm 0.02 & 1.00 \pm 0.01 & 0.97 \pm
0.03 & 0.99 \pm 0.01 \\ 
c_1 & -1.85 \pm 1.34 & -5.14 \pm 10.46 & 0.53 \pm
0.40 & -1.76 \pm 0.14 & -1.33 \pm 0.46 & -1.73 \pm 0.03 \\
c_2 & 5.09 \pm
9.46 & 34.05 \pm 93.82 & -15.14 \pm 2.87 & 1.60 \pm 0.84 & -1.40 \pm 2.60
& 0.48 \pm 0.08 \\ 
c_3 & -14.99 \pm 31.38 & -157.82 \pm 428.42 & 58.03
\pm 9.82 & -2.22 \pm 1.97 & 5.68 \pm 6.05 & \\ 
c_4 & 17.1 \pm 49.72 &
405.88 \pm 1059.89 & -105.52 \pm 15.92 & 0.07 \pm 1.62 & -7.18 \pm 5.00 &\\ 
c_5 & -7.82 \pm 30.32 & -558.73 \pm 1353.59 & 71.97 \pm 9.92 & & &\\ 
c_6 & & 318.7\pm700.1 & & & &\\
\hline
\frac{\chi^2}{dof} & 2.46 & 1.75 & 2.32 & 5.97 & 7.47 & 8.03\\
dof & 12 & 7 & 4 & 5 & 5 & 4\\
\hline
\end{array}
\]
\end{scriptsize}

\caption{Interpolation next-best fit parameters for $Z_P$.
}
\label{tab:zfit2}
\end{table}

\bibliographystyle{unsrt}

\end{document}